\documentclass{ws-p8-50x6-00}

\begin{document}

\title{Trinification from superstring toward MSSM}

\author{Jihn E. Kim}

\address{School of Physics and Center for Theoretical Physics,\\
Seoul National University,
Seoul 151-747, Korea\\E-mail: jekim@phyp.snu.ac.kr}

\maketitle\abstracts{
In this talk, I present a family unification in $Z_3$
orbifolded $E_8\times E_8^\prime$ heterotic strings. It
is argued that trinification is a plausible candidate toward
supersymmetric standard model at low energy.}

\def\one{\bf 1}
\def\three{\bf 3}
\def\threeb{\bf{\bar 3}}

\section{Introduction}

Even though the standard model(SM) is very successful, there
exists the so-far unsolved family problem that there are 3 sets
of fifteen chiral fields. $\lq\lq$Is 3 a very fundamental
number in the universe?" We try to investigate this problem.

A grand unification(GUT) in $SO(10)$ is very promising, but it has
only one family. To have three families, one has to repeat the
representations, which we want to avoid. In this respect a family
symmetry such as $SO(3)$ or $SU(3)$ has been considered. But these
extra family group will face the problem of Goldstone bosons or
gauge unification. This is the reason to go beyond $SO(10)$, i.e.
to $SO(4n+2)$ with one spinor representation toward the unification
of flavors~\cite{kim80}. Most part of $SO(4n+2)$ can be studied in
$SU(2n+1)$. For example, SU(7) with the fermion representation,
$\Psi\oplus\Psi^A\oplus \Psi_{AB}\oplus \Psi^{ABC}$ 
contains 64 components of
the $SO(14)$ spinor {\bf 64}. But a naive breaking of $SO(14)$ down to
$SO(10)$ leads no chiral fermions. One must {\it twist} the gauge group
to obtain chiral fermions~\cite{kim80}. This model, however, lacks the
third quark families and is not phenomenologically successful.
In addition, there are extra particles not present in the SM. In many
models with twisting, extra particles are unavoidable.
However, this road of attractive grand unification of flavor has
not been considered any more since 1984, 
due to the possibility of understanding
the family structure in the $E_8\times E_8^\prime$ heterotic
string model~\cite{heterotic}.

One crucial thing needed for the flavor grand unification is that the gauge
group should be big enough. Another thing is that the fermion representation
should be anomaly free. It can be a reason to exclude the $SU(N)$ gauge
groups since to cancel anomalies the $SU(N)$ representations 
should be matched miraculously.
In this respect, $SO(4n+2)$ allowing complex representations attracted
a great deal of attention. Typically one assumes one spinor representation
in $SO(4n+2)$. But, here introducing $\lq$one' spinor does not have a strong
rationale for the one. In this respect, we note that the gauge bosons has
a fixed representation, i.e. the adjoint representation. So, it may be
reasonable to relate fermions to the adjoint representation of the
gauge group. To introduce fermions, it is better to have supersymmetry.
Indeed, this road is exactly what the higher dimensional
gauge theories take, and in particular in the superstring models.

\section{Need for HESSNA}

Superstring models are written in 10 dimensions(10D).
A simple dimensional reduction down to 4D would not lead to chiral
fermions. One has to twist the gauge group to obtain chiral families
as in the previous example of $SO(14)$~\cite{kim80}. So, twisting the gauge
group is necessary. Since we have to hide 6 internal spaces through
compactification, there is a possibility for twisting the internal
space. There are two well-known compactifications achieving these
goals, the Calabi-Yau space compactification~\cite{chsw} and
the orbifold compactification~\cite{dhvw}. Among these the orbifold is
simpler and easy to understand geometrically. So, the orbifold
compactifications are most extensively studied.

The 4D string models were constructed in orbifold construction and
in fermionic construction. The first standard-like models were constructed in
orbifold compactifications~\cite{iknq,imnq}. On the other hand,
the flipped $SU(5)$~\cite{barr} was constructed in the fermionic
construction~\cite{ellis}. These 4D string models can be considered
as a 4D theory, not coming from 10D. But it is tempting to speculate
that 4D models are the remnants of compactification of the 10D string,
in which case the orbifold has the merit of geometrical interpretation,
compared to the fermionic construction.

The simplest orbifold is to consider three two-dimensional tori
$6D=2D+2D+2D$ with a further identification of the point group.
Among many $Z_N$ orbifold models, $Z_3$ orbifolds with
$N=1$ supersymmetry are especially fascinating
because the multiplicity of matter fermions come in multiples of 3.
This may be the reason that the family number is 3.

\begin{figure}[t]
\begin{center}
\epsfxsize=10pc
\epsfbox{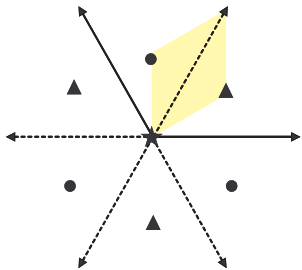}
\end{center}
\caption{ The fundamental
region of $Z_3$ orbifold is shaded. \label{fig:Z3}}
\end{figure}

The $Z_3$ orbifold identifies the points related by $120^o$ rotation,
and hence there are three fixed points in the fundamental region as
shown in Fig. 1. Since we must consider the direct product of three tori
there are 27 fixed points. This geometrical twisting can be also
acompanied in the twisting of the gauge group. The frequently discussed
example is the shift vector $v=\frac13 (1~1~2~0~0~0~0~0~)(0~\cdots)$ which
gives the gauge group $E_6\times SU(3)\times E_8^\prime$. Since $E_6$
contains the spinor representation of $SO(10)$ it attracted a great
deal of attention. The reason that the fundamental representations of
$E_{6,7,8}$ contain the $SO(10)$ spinor is the main phenomenological
reason favoring the $E_8\times E_8^\prime$ heterotic string. In this
talk also, we focus on the $E_8\times E_8^\prime$ heterotic string.

However, the symmetry breaking of $E_6$, $SO(10)$ and $SU(5)$ down to the
standard model requires an adjoint representation for the GUT Higgs field(s).
But in orbifold compactifications it is very difficult to obtain the
adjoint representation\footnote{For a high level Kac-Moody
algebra, it is known that the adjoint representation is possible~\cite{tye}.}.
This is the reason that the flipped $SU(5)$ GUT attracted
so much attention~\cite{ellis}. Because of this difficulty of GUT
symmetry breaking, the direct derivation of the SM gauge group with
reasonable fermionic spectrum attracted a great deal of
attention~\cite{iknq,imnq}, and are called standard-like models.
These standard-{\it like} models pursue the following properties:
\begin{itemize}
\item The gauge group is $SU(3)\times SU(2)\times U(1)^n$.
\item There are three families.
\item In some cases, there are Higgs doublets but no color
triplets~\cite{iknq}.  This is the doublet-triplet splitting.
\end{itemize}
In this talk, we try to delete $like$ from standard-{\it like}.

There are two reasons that the standard-like models are not phenomenologically
successful. One is the $\sin^2\theta_W$ problem in that most of these
standard-like models give the string value of $\sin^2\theta_W$ too small
compared to $\frac38$~\cite{kim03}. Another problem is that there are
too many Higgs doublets appearing in
the spectrum. Usually, the minimum number of the Higgs
doublets is six, 3 from $Z_3$ and 2 from $H_{u,d}$ for
the anomaly cancellation.
If the SM is embeeded in a simple GUT group, the bare value of the
$\sin^2\theta_W$ is
\begin{equation}
\sin^2\theta_W^0=\frac{{\rm Tr~}T_3^2}{{\rm Tr~}Q_{em}^2}
\end{equation}
where $T_3$ is the 3rd component of weak iso-spin and
$Q_{em}$ is the electromagnetic charge. In general, there appear many
charged electroweak singlet fields and hence $\sin^2\theta_W$ can be
much smaller than $\frac38$. This is in gross contradiction with
the LEP measurement  of the seemingly unified gauge coupling constant
at $2\times 10^{16}$~GeV.\footnote{To cure this problem, the so-called
optical unification was suggested~\cite{giedt}.}
The basic reason of the $\sin^2\theta_W$ problem is that the
electroweak hypercharge $Y$ is leaked to uncontrollably many $U(1)$'s.
A GUT is permissible in orbifold compactification,
but it is difficult to obtain an
adjoint representation. This has led to the consideration of the
flipped $SU(5)$, i.e. $SU(5)\times U(1)$, but in the flipped
$SU(5)$ there is also the problem of the leakage of the
electroweak hypercharge to the extra $U(1)$.
Therefore, we suggest GUT groups with the
following property~\cite{kim03}:
\begin{itemize}
\item HESSNA = Hypercharge is Embedded in Semi-Simple group
with No need for Adjoint representation.
\end{itemize}
Most probably, the QCD $SU(3)$ is already separated out, except
in Pati-Salam type GUT~\cite{hdkim}. The simplest HESSNA is the
trinification group $SU(3)_1\times SU(3)_W\times SU(3)_c$
with the fermion representation
\begin{equation}\label{tri}
(\threeb,\three,\one)+(\one,\threeb,\three)+(\three,\one,\threeb).
\end{equation}
At non-string level the trinification has been extensively
studied~\cite{trif},
but here the only requirement is the anomaly cancellation and
phenomenological massage. But in the string trinification the
theory is very restrictive. It is dictated from string theory.
Breaking of $SU(3)_1\times SU(3)_W\times SU(3)_c$ down to the SM is
achieved by giving VEV's to two fields in the spectrum (\ref{tri}).
The electroweak hypercharge is represented as
\begin{equation}
Y=-\frac12(-2I_1+Y_1+Y_W)
\end{equation}
with the subscripts denoting the respective $SU(3)$ groups.

\section{$Z_N$ embedding in $E_8$}

Therefore, we looked for $SU(3)^3$ trinification groups from
$E_8\times E_8^\prime$ heterotic string. We used the Dynkim
diagram technique to find out the gauge groups, by embedding
$Z_N$ in $E_8$. This method was
originally devised by Kac and Peterson~\cite{kac}, starting
from the extended Dynkin diagram(Fig. 2) of $E_8$. The rank-8
$E_8$ has eight simple roots. The extended diagram has
the ninth root $\alpha_0$, satisfying $\alpha_0\cdot
\alpha_i=0$ if $i\ne 1$, and $\alpha_0\cdot\alpha_1=-1$.
The Dynkin basis $\{\gamma_i;\ i=1,2,\cdots,8\}$ is defined
to satisfy $\gamma_i\cdot\alpha_j=\delta_{ij}$. In this Dynkin
basis a shift vector $V$ is expanded. The embedding of $Z_N$
shift vector is $V=\frac{1}{N}\sum_i s_i\gamma_i$ with
\begin{equation}
\sum_{i=1}^{8}n_i\gamma_i=0\ \ {\rm mod.~} N,\ \ {\rm for~}
s_i\ge 0.
\end{equation}
Now for $s_i\ne 0$, remove that circle in the extended Dynkin
diagram, which gives the surviving group with the $U(1)$'s
added to make up rank 8. But the generalization with more than
one shift vector is necessary to wrap the tori with Wilson
lines~\cite{inq}. Recently, we resolved this problem of adding
more shift vectors~\cite{chk}.\footnote{At field theory level,
a similar study has been done~\cite{hebecker}.}
This Dynkin diagram technique is a great help in finding out
HESSNA, not counting the same model several times.
Of course, it is better than trying every possible $V$ and
$a$'s, since we can figure out the gauge group from the
beginning. Indeed, this method was the guiding idea finding
out early string trinifications~\cite{octa,ck03}.

\begin{figure}[h]
\begin{center}
\epsfxsize=20pc \epsfbox{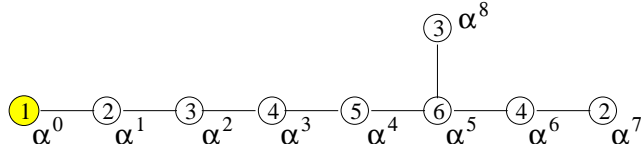}
\end{center}
\caption{Extended Dynkin diagram of $\widehat{E_8}$ group. The
numbers in the circle are the Coxeter label $n_i$ of the
corresponding simple roots. $\alpha_0$ is the extended one.}
\end{figure}

\section{A mass matrix ansatz for MSSM}

Now, let us show one example how one can obtain a MSSM.
For $Z_3$ orbifolds, the bulk(untwisted) matter fields have
multiplcity 3. The fields located at the fixed
points have multiplicity 27. One Wilson line models
give the multiplicity 9 at a fixed sector. Two
Wilson line models give the multiplicity 3 at a fixed
sector. Therefore, to construct a model with a reasonable
spectrum it is plausible to consider two Wilson line models.
It may be tempting to consider three Wilson line models
which distinguish 27 fixed points. However, with
multiplicity=1 three families
are not guaranteed at the outset. Therefore, two
Wilson line models are the best at the moment. The Wilson
lines($a_i\ (i=1,3)$) are shifts. In applying the Dynkin
diagram technique, it is the same as the shift vector $V$,
but we must satisfy the modular invariance conditions:
\begin{eqnarray}
&V^2=\frac23\cdot({\rm integer}),\ \ a_i^2=\frac23
\cdot({\rm integer})\nonumber\\
&V\cdot a_i=\frac13\cdot({\rm integer})\\
&a_i\cdot a_j=\frac13\cdot({\rm integer\ \ for})\ i\ne j
\nonumber
\end{eqnarray}

The trinification spectrum comes in three different
types of representations which we call the {\it lepton humor}$(
\threeb,\three,\one)$ {\it quark humor}$(\one,\threeb,\three)$,
and {\it anti-quark humor}$(\three,\one,\threeb)$, respectively.
The trinification spectrum (\ref{tri}) is very similar to {\bf 27}
of $E_6$, as far as the spectrum is concerned. However, for the
GUT symmetry breaking, the trinification is much better. The
trinification spectrum contains two neutral components($N_{10}$
and $N_5$ in the lepton-humor sector) the VEV's of which can
break $SU(3)^3$ down to the SM gauge group.

For a definite presentation of our argument, let us take
a specific two Wilson-line $Z_3$ trinification model~\cite{cchk},
\begin{eqnarray}
& V=(0~0~0~0~0~\frac13~\frac13~\frac23)(0~0~0~0~0~0~0~0)
\nonumber\\
& a_1=(\frac13~\frac13~\frac13~0~0~\frac13~\frac23~0)
(0~0~0~0~0~0~0~\frac23)\\
& a_3=(0~0~0~0~0~\frac13~0~\frac13)
(0~0~0~0~\frac13~\frac13~\frac13~\frac13)
\nonumber
\end{eqnarray}
which gives the $N=1$ supersymmetry with the following gauge group
\begin{equation}
SU(3)^3\times U(1)^2\times [SO(8)\times SU(3)\times U(1)^2]_h.
\end{equation}
The chiral superfields are shown in Table 1. We can remove
vectorlike representations as has been the practice in GUT's. It
is possible by giving VEV's to the gauge group singlets. For a string
calculation one should follow the Yukawa coupling derivation
given in~\cite{hv87}, but here we simply adopt the old GUT
procedure since our presentation is just an idea toward MSSM.
Then, we obtain the trinification spectrum from T0, and in addition
more trinification-like fields,
\begin{eqnarray}
& (\threeb,\three,\one)(\one,\one)+(\one,\threeb,\three)(\one,\one)
+(\three,\one,\threeb)(\one,\one)\nonumber\\
&+(\threeb,\three,\one)(\one,\one)+(\three,\one,\one)(\one,\three)
+(\one,\threeb,\one)(\one,\threeb).
\end{eqnarray}

\begin{table}[t]
\caption{The massless spectrum of the orbifold. The 3rd column
denotes the multiplicity.}
\vspace{0.2cm}
\begin{center}
\footnotesize
\begin{tabular}{c|c|c|c}
\hline
sector & twist & mul. & fields \\
\hline
U & & 3 & $(\threeb,\three,\one)(\bf 1,1)$ \\
T0 & $V$ & 9 & $(\one,1,1)(\bf 1,1)$ \\
  &  & 3 & $(\threeb,\three, 1)(\one,1)+
(\three,\one,\threeb)(\one,1)+(\bf 1,\threeb,\three)(\one,1)$ \\
T1 & $V+a_1$ & 3 & $(\one,\three,1)(\one,1)+(\three,1,1)
(\one,1)+(\one,1 ,\three)(\one,1)$ \\
T2 & $V-a_1$ & 3 & $(\one,\threeb,1,)(\one,1)+(\threeb,1,1)
(\one,1)+(\one,1,\threeb)(\one,1)$ \\
T3 & $V+a_3$ & 9 & $(\one,1,1)(\one,1)$ \\
  & & 3 & $(\one,1,1)(\one,\threeb)+(\one,1,1)(\one,\three)$\\
 & & & $+
(\one,1,1)(\one,1)+(\one,1,1)(\bf 8,1)$ \\
T4 & $V-a_3$  & 9 & $(\bf 1,1,1)(\bf 1,1)$ \\
  & & 3 & $(\bf 1,1,1)(\bf 1,3)+(\bf 1,1,1)(\bf 1,\threeb)$\\
& & &
$+(\bf 1,1,1)(\bf 1,1)+(\bf 1,1,1)(\bf 8,1)$ \\
T5 & $V+a_1+a_3$ & 3 & $(\three,1,1)(\bf 1,3)$ \\
T6 & $V+a_1-a_3$ & 3 & $(\bf 1,\three,1)(\bf 1,1)+
(\three,1,1)(\bf 1,1)+(\bf 1,1,\three)(\bf 1,1)$ \\
T7 & $V-a_1+a_3$ & 3 & $(\bf 1,\threeb,1)(\bf 1,1)+(\threeb,1,1)
(\bf 1,1)+(\bf 1,1,\threeb)(\bf 1,1)$ \\
T8 & $V-a_1-a_3$ & 3 & $(\bf 1,\threeb,1)(\bf 1,\threeb)$ \\
\hline
\end{tabular}
\end{center}
\end{table}

It is known that a vectorlike lepton-humor is needed toward
neutrino masses and reasonable symmetry breaking
pattern~\cite{veclep}. We have a lepton humor in U, but does not
have its antiparticles. So, the gauge group must be identified
toward this purpose. We can identify $SU(3)_1$ from $E_8$ and
$\overline{SU(3)}_h$ from $E_8^\prime$ by the linkage field
in T5, and obtain anti-lepton-humor from T8. Because there appear
many Higgs doublets in the resulting spectrum, we take an ansatz
that determinant of the Higgsino mass matrix vanish~\cite{cchk},
\begin{equation}
{\rm Det.}~M_{\tilde H}=0.
\end{equation}
Then, we obtain only one pair of Higgs doublets at low energy,
realizing the MSSM spectrum. This ansatz can be dictated from
dynamics at high energy, such as the small instanton effects.
The relevan instanton absorbs the vectorlike representations
of Higgsinos as shown in Fig. 3. Then in this scheme
it is possible to remove vectorlike color triplets~\cite{cchk}.
Even though the specific model we discuss supports our ansatz, it
does not give a bare value of $\sin^2\theta_W^0=\frac38$. So,
we may not take it as a fully satisfactory model. It will be
seen whether a model realizing the Higgsino mass matrix ansatz
with $\sin^2\theta_W^0=\frac38$ is present in $Z_3$ orbifold
models.

\begin{figure}[t]
\begin{center}
\epsfxsize=10pc \epsfbox{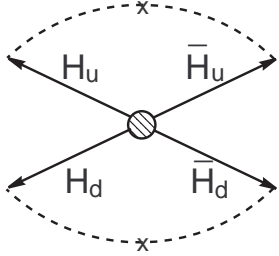}
\end{center}
\caption{A possible instanton interaction. All the Higgsino
pairs should be included.}
\end{figure}

\section{Conclusion}

In this talk, I showed a road toward the construction of MSSM through
the $Z_3$ orbifold compactification of the $E_8\times E_8^\prime$
heterotic string. The $Z_3$ is chosen to interpret three
families. In standard-like models, there are too many Higgs doublets
present. The problems of the extra $U(1)$'s and too many Higgs
doublets are the obstacles for a bare weak mixing angle being
$\frac38$. Therefore, gauge groups with no need
of adjoint representation are proposed as GUT groups for an easy
pattern of symmetry breaking. In particular, a trinification
(\ref{tri}) has been suggested toward a GUT group between the
string scale and the electroweak scale.
To obtain a MSSM from HESSNA, one should allow
only one pair of Higgs doublets. We suggested a short distance
dynamics toward realization of one pair of Higgs doublets
through the ansatz, Det.~$M_{\tilde H}=0$.
To obtain a bare $\sin^2\theta_W^0=\frac38$, it is better to
obtain a trinification spectrum $6({\bf 27})\oplus
3({\bf \overline{27}})$, or $3({\bf 27})\oplus
$3(lepton-humor + anti-lepton-humor),
where {\bf 27} is the trinification spectrum.
But, we have not obtained such a $Z_3$ model yet.

\section*{Acknowledgments}
This work is supported in part by the KOSEF
ABRL Grant No. R14-2003-012-01001-0,
the BK21 program of Ministry of Education, and Korea
Research Foundation Grant No. KRF-PBRG-2002-070-C00022.


\end{document}